\documentclass[preprint2]{aastex}

\newcommand{\NNN}{48 }
\newcommand{\NNc}{432 }
\newcommand{\NNpoor}{61}

\newcommand{\Ha}{H${\alpha}$ }
\newcommand{\Hb}{H${\beta}$ }
\newcommand{\OIII}{[O${\rm III}]\lambda$5007 }
\newcommand{\NII}{[N${\rm II}]\lambda$6584 }
\newcommand{\de}{$^\circ$ }
\newcommand{\ddd}{D4000~}

\newcommand{\kms}{km\,s$^{-1}\,$}

\newcommand{\Msunyr}{$M_{\odot}\, yr^{-1}\,$}
\newcommand{\Msunyrkpc}{$M_{\odot}\, yr^{-1}\, kpc^{-2}\,$}
\newcommand{\aFe}{[$\alpha$/Fe]~}

\shorttitle{Star-Formation Driven Biconical Outflows}
\shortauthors{Bizyaev et al.}

\begin{document}

\title{SDSS IV MaNGA - Star-Formation Driven Biconical Outflows in the Local
Universe}

\author{
Dmitry Bizyaev  \altaffilmark{1,2,3}, 
Yan-Mei Chen  \altaffilmark{4,5}, 
Yong Shi  \altaffilmark{4,5}, 
Rogemar A. Riffel  \altaffilmark{6,7,8}, 
Rogerio Riffel  \altaffilmark{6,9}, 
Aleksandar M. Diamond-Stanic  \altaffilmark{10}, \&
Namrata Roy \altaffilmark{11}
}

\altaffiltext{1}{Apache Point Observatory and New Mexico State University, Sunspot, NM, 88349, USA}
\altaffiltext{2}{Sternberg Astronomical Institute, Moscow State University, Moscow, Russia}
\altaffiltext{3}{Special Astrophysical Observatory of the Russian AS, 369167, Nizhnij Arkhyz, Russia}
\altaffiltext{4}{School of Astronomy and Space Science, Nanjing University, Nanjing 210093, China}
\altaffiltext{5}{Key Laboratory of Modern Astronomy and Astrophysics (Nanjing University), Ministry of Education, Nanjing 210093, China}
\altaffiltext{6}{Laborat\'orio Interinstitucional de e-Astronomia, 77 Rua General Jos\'e Cristino, 20921-400, Rio de Janeiro, Brazil}
\altaffiltext{7}{Universidade Federal de Santa Maria, Departamento de F\'isica, Centro 
de Ci\^encias Naturais e Exatas, 97105-900, Santa Maria, RS, Brazil}
\altaffiltext{8}{Department of Physics \& Astronomy, Johns Hopkins University, Bloomberg
Center, 3400 N. Charles St, Baltimore, MD 21218, USA}
\altaffiltext{9}{Instituto de F\'isica, Universidade Federal do Rio Grande do Sul, Campus do Vale, 91501-970 Porto Alegre, Brazil}
\altaffiltext{10}{Bates College, Department of Physics \& Astronomy, 44 Campus Ave, Carnegie Science Hall, Lewiston, ME 04240, USA}
\altaffiltext{11}{University of California Santa Cruz, 1156 High Street, Santa Cruz, CA 95064, USA}

\begin{abstract}

We present a sample of 48 nearby galaxies with central, biconical outflows identified   
by the Mapping Nearby Galaxies at APO (MaNGA) survey.
All considered galaxies have star formation driven bi-conical central outflows (SFB), with no signs of AGN. 
We find that the SFB outflows require high central concentration of the 
star formation rate. This increases the gas velocity dispersion over the equilibrium limit
and helps maintain the gas outflows.
The central starbursts increase 
the metallicity, extinction, and the \aFe ratio in the gas. 
Significant amount of young stellar population at the centers suggests that the SFBs are 
associated with the formation of young bulges in galaxies. 
More than 70\% of SFB galaxies are group members or have companions 
with no prominent interaction, or show asymmetry of external isophotes. In 15\% SFB cases 
stars and gas rotate in the opposite directions, which 
points at the gas infall from satellites as the primary reason for triggering the SFB phenomena. 

\end{abstract}


\section{Introduction}

Biconical outflow of gas from the centers of galaxies is a typical manifestation of 
processes powered by active galactic nuclei (AGN). 
Gas outflows can be also powered by intensive star formation and driven by 
supernovae and stellar winds from young stellar population \citep{strickland07}. They are found to 
be ubiquitous in galaxies at low \citep{veilleux05,chen10} and high redshifts \citep{rubin14,davies18}.
Intensive outflows driven by star formation bursts in galaxies without AGNs found in  
nearby starburst galaxies were dubbed superwinds \citep{superwinds0,superwinds,heckman15}. 

First time the superwinds were detected in M~82 \citep{lynds63,burbidge64} as a gas outflow from the center.
In most cases of the superwinds the warm gas escapes the galaxies in all directions \citep{superwinds}, whereas
just a few local galaxies, e.g. M~82 and NGC~253, demonstrate
biconical star formation driven superwinds stemmed from the centers \citep{heckman90,lehnert99}. 
Large mass of warm gas is ejected in superwinds 
 \citep{heckman02,chisholm15} at a high rate (10-20 \Msunyr)
to the distance of dozen kpc \citep{veilleux03,strickland07} with hundred \citep{shopbell98,veilleux05} 
to thousand \kms speed \citep{heckman00}
when the star formation surface density exceeds a threshold for the superwinds of 0.1 \Msunyrkpc
\citep{heckman02}.
Numerical simulations predict the ubiquity of the central bi-conical outflows
\citep{ttmt98,fielding17,schneider18}.

Galactic winds help inhibit star formation and remove baryonic matter
from star formation sites, which affects the chemical evolution in galaxies, shape 
integral scaling relations \citep{veilleux05,fielding17} and enrich the intergalactic gas
media. In turn, the outflows may remove the gas from the galaxies only temporary
\citep{oppenheimer06,oppenheimer10,leroy15}, which makes them an important 
component of the extragalactic gas supply circulation process. The centrally concentrated, star formation
driven outflow provides a simplified case with clear central localization.
Exploring the reasons for starting and maintaining the central, biconical outflow 
helps better understand processes of gas exchange between the high-altitude intergalactic medium 
and low-altitude sites of active star formation. 

Statistical power of large integral field spectroscopic surveys allows us to look for more examples of the 
centrally concentrated outflows in the local Universe \citep{roche15,ho16,gallagher19} and 
to determine the key factors that play role
in the life cycle of the outflows \citep{veilleux05,zhang18}.
In this paper we focus on observational manifestation and properties of the central, 
biconical star formation driven outflows identified in a large sample of galaxies. We explain our selection
procedure and the outflow galaxies sample formation criteria in \S2. We also create
a sample of regular galaxies, without noticeable outflows, which we use for comparison purposes. 
The observational properties of the galaxies in the samples are examined in \S3. 
In \S4 we explore the differences between the 
galaxies with detected outflows and regular galaxies, and discuss unique features of 
the galaxies with the central, star formation driven outflows.

\section{The Sample of Star Formation Driven Bicones from MaNGA observations}

We look for the star formation bicones (SFB) in a large sample of relatively nearby galaxies 
using results from Mapping Nearby Galaxies at APO (MaNGA, \citet{manga}) survey performed at a dedicated 2.5-m telescope \citep{25m}.  
MaNGA, a part of the Sloan Digital Sky Survey \citep{sdss_old,sdss}, 
is a multiple \citep{drory15,law16} Integral Field Unit survey of several thousand local galaxies (median redshift $\sim$ 0.03 \citep{yan16a}
with spectral resolution of 2000 and the 3,600-10,300\AA~ wavelength coverage \citep{boss_spectrographs}. 
The survey's target selection \citep{wake17} provides a roughly uniform stellar mass distribution 
for MaNGA galaxies and allows us to obtain kiloparsec-scale
spatial resolution maps of the stellar and ionized gas kinematics \citep{law15}.


We started with the MaNGA "product launch" MPL-6 \citep{law16} which released 
4857 objects, estimated the inclination
from the ellipticity of SDSS images as in \citet{chen10}, and selected a sample of 
1589 galaxies
with the inclination of 60\de or higher. Next, we selected only the galaxies in which
the central and eight adjacent spaxels 
fall onto the star formation region on the BPT diagram \citep{bpt}. 
Only the spaxels with the signal-to-noise ratio of 3 and greater in all significant for this study
emission lines (\Ha, \Hb, \NII, and \OIII) were considered. 
As an independent check, we check 
the galaxy BPT classification made in the MPA-JHU catalog \citep{mpa-jhu}, which uses SDSS 
spectroscopy data, independent of MaNGA. 
All SFB sample and control galaxies were confirmed to have star-forming 
centers in the MPA-JHU catalog. 

\begin{figure}
\epsscale{1.0}
\plotone{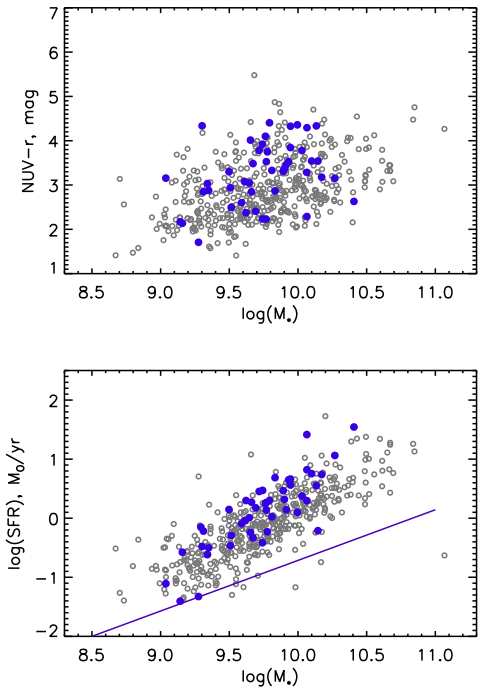}
\caption{
The MaNGA galaxies with (blue bullets, SFB sample, see text) and without (grey circles, control sample, see text) 
the star forming bi-conical structures on the SFR - Stellar Mass diagrams. The stellar mass is taken from the 
NASA-Sloan Atlas \citep{nsa}.
Top: the near-ultraviolet - red magnitude difference estimated by the NASA-Sloan Atlas.
Bottom: the integrated star formation rate in the galaxies from the MaNGA data, see \S 3.1.
The solid line designates the demarkation between the star forming and green valley galaxies
according to \citet{chang15}
\label{f0}}
\end{figure}

As one more verification,
we place all considered galaxies on the SFR-Stellar Mass diagram, see Figure \ref{f0} and ensure that 
the objects with and without the SFB occupy the same region, which corresponds
to the star formation, blue cloud. 
We inspect two-dimensional maps 
of the \Ha and \OIII equivalent width (EW) and maps of the gas velocity dispersion 
in all galaxies. 
Note that we identify the extra-planar ionized gas structures similar to those reported
by \citet{cheung16,roy18,riffel19}, who notice ionized gas regions with enhanced \Ha and \OIII emission 
extended along the minor axes in a sample of early-type galaxies. 
We split the galaxies by three groups:
first one with the biconical structures aligned along the minor axis in the 
emission lines; second group without these features, and the third group with uncertain 
or disturbed maps of the  \Ha and \OIII EW.  The 
latter group includes also  galaxies with perturbed emission line fields, for which we 
cannot claim a clearly identified SFB. The gas velocity dispersion fields were
inspected for the galaxies with prominent bi-conical structures. All these galaxies demonstrate
enhanced velocity dispersion with respect to the galactic periphery. 

\begin{figure}
\epsscale{1.20}
\plotone{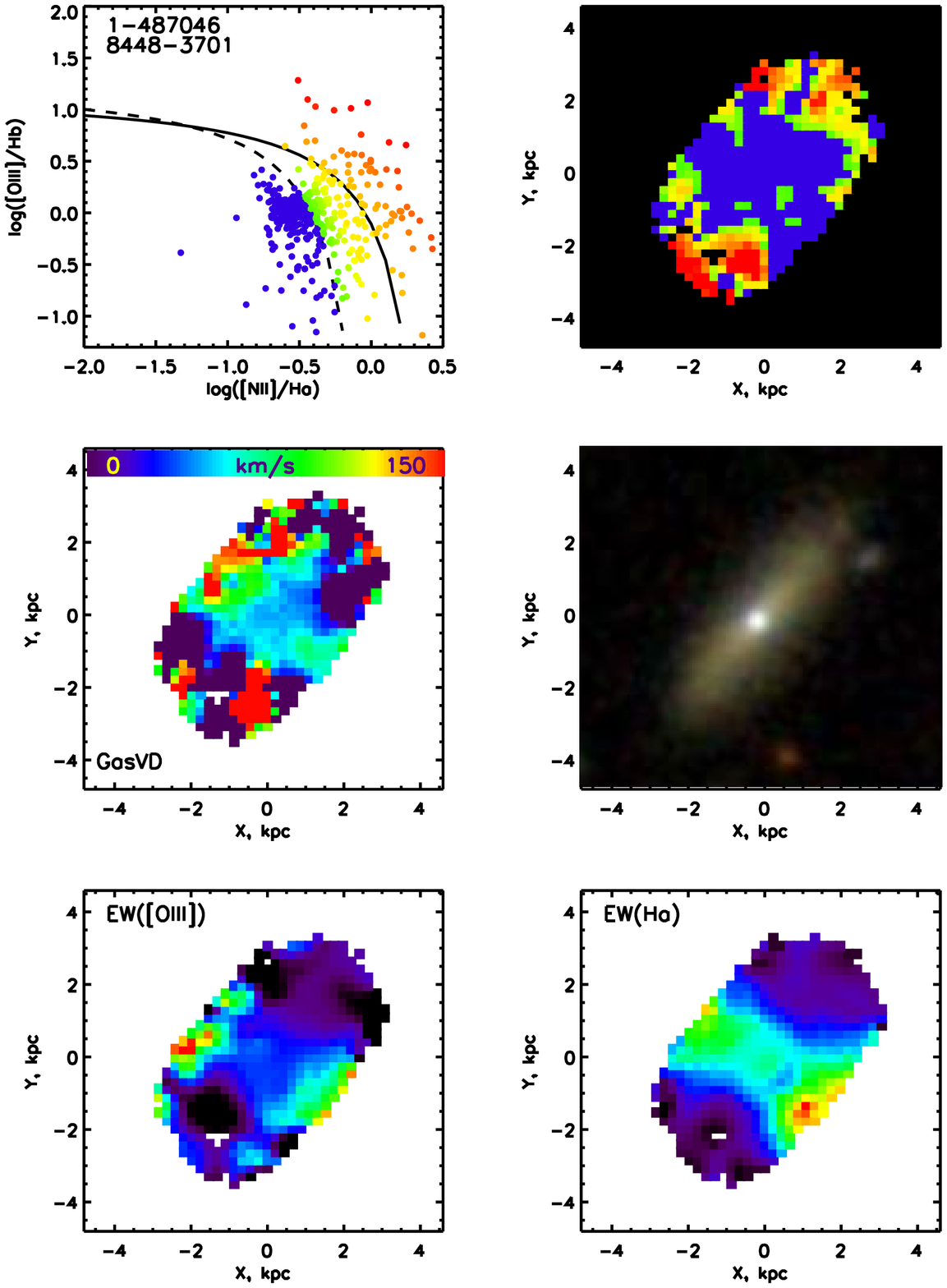}
\caption{
    One of our galaxies, 8448-3701.
    Top left: a BPT diagram for MaNGA spaxels with signal-to-noise ratio S/N $>$ 5 in the \Ha flux. 
   The colors designate the star formation (blue), AGN/LINER (red) and composite (yellow/green) spaxels.   
   The demarkation curves by \citet{bpt_ke01,bpt_ka03} separate the star formation, AGN and 
   composite regions. 
   Top right panel: the distribution of the color coded BPT spaxels over the galaxy.
   Middle left panel: the ionized gas velocity dispersion map. 
   Middle right panel: SDSS color image of the galaxy.  
   Bottom left panel: the [OIII] equivalent width map of the galaxy. The emission distribution reveals the 
   biconical outflow along the minor axis of the galaxy associated with high gas velocity dispersion.
   Bottom right panel: the \Ha equivalent width map of the galaxy. 
\label{f1}}
\end{figure}

Our inspection of the gas and stellar kinematic maps revealed 7 SFB galaxies
with counter-rotating gas and stars, which means close to 180\de difference in the 
spin vectors between them in the case of edge-on galaxies. We also identified
several cases of perpendicular rotation between gas and stars, and rejected such objects from the samples 
as polar ring galaxies. Any interacting or warped objects, as well as members of 
tight groups or pairs, were removed from all samples. 
We tailor the control sample of galaxies such that
the galaxies with and without SFB have very similar distribution
by their distances, stellar mass, and Sersic index. 
The first group forms the target sample of \NNN galaxies with biconical 
outflow (SFB). From the second group we selected the galaxies in the radial velocity range
and stellar mass (according to the NSA catalog)  similar to the target sample (5000 and 14000 \kms), and
thus formed the control sample of \NNc objects, which we use for comparison purposes. 
Thus, the control galaxies are similar to the SFB objects, but do not show enhanced
emission along the minor axis.
We removed the third group of \NNpoor\, uncertain galaxies from the further consideration. 
An example of our galaxies with Star Formation Bicones is shown in Figure \ref{f1}.
The MaNGA galaxy 8448-3701 shown in Figure \ref{f1} has star-forming only spaxels in the 
central area. The optical SDSS image looks rather regular, but the emission line maps
demonstrate clear structures aligned along the minor axis of the galaxy, in which the
gas velocity dispersion is also higher than in the rest of the galaxy.

\section{Observational Properties of the Galaxies with Star Formation Driven Bicones}

\subsection{The Star Formation Rate and the Extinction}

\begin{figure}
\epsscale{1.00}
\plotone{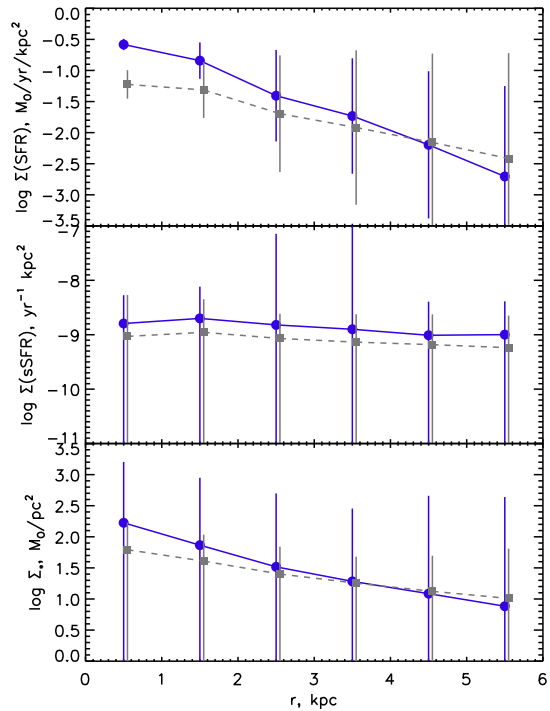}
\caption{The median values for the star formation rate surface density (top),
specific star formation rate surface density (middle), and stellar surface density (bottom) for the 
SFB galaxies (blue symbols and solid lines) and control sample (grey symbols and dashed lines).
The error bars reflect the 1-$\sigma$ standard deviation of the objects in each bin.
\label{f2a}}
\end{figure}

The galactic effective radii, Sersic index and estimated stellar masses are adopted from the 
NASA-Sloan Atlas \citep{nsa} (NSA\footnote{http://nsatlas.org}).
Although the SFB galaxies have similar stellar mass and distance distribution, 
some of their features differ significantly from those in the control sample. 
We compare the median radial profiles
of the star formation rate, specific star formation rate, and stellar surface density for the 
SFB and control sample in Figure \ref{f2a} and find that only the star formation surface density 
is significantly different in the SFB galaxies, and only in their central regions. 

The internal extinction $A_V$ in the galaxies is estimated through the Balmer decrement 
via \Ha/\Hb ratio with the assumption of Case B recombination \citep{osterbrock06} extinction-free ratio of 2.86
and Cardelli et al extinction law \citep{cardelli89}. 

Each galaxy is split by elliptical annuli 1 kpc wide if the inclination was less than 
80\de, or by zones of the same width if the inclination was higher.
The reported $A_V$ is an average value in each zone or annulus. 
Only the spaxels with no bad reduction flags and S/N $>$ 3 in the \Ha and \Hb flux are 
used for the $A_V$ calculations. The same elliptical annuli or zones were used to 
estimate the star formation rate from the extinction corrected \Ha surface
brightness \citep{martin01}.
The \Ha surface density concentration is estimated as the ratio of \Ha luminosity 
within and outside the central 1 kpc circle.

\begin{figure}
\epsscale{1.00}
\plotone{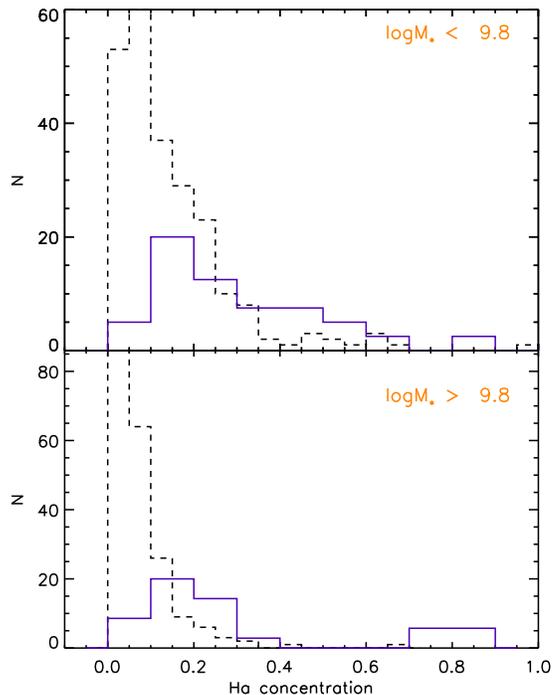}
\caption{The central concentration of \Ha surface brightness for the 
SFB (blue solid) and control (black dashed) galaxies in the low (top) and high (bottom) mass galaxies
\label{f2}}
\end{figure}

We split all galaxies into two groups with low and high stellar masses
(corresponding to log($M_*/M_{\odot}$) under or over 9.8 dex.)
The central surface density of star formation is higher in the SFB galaxies of all stellar mass groups.
The \Ha luminosity concentration 
is also prominently higher in the SFB galaxies with respect to the regular ones, see Figure \ref{f2}.
These trends suggest that the star formation is the primary mechanism that powers up the 
gas outflow, in agreement with \citet{heckman02}. 


The galactic extinction $A_V$ in the SFB galaxies is higher than that in the control sample, on average, especially 
in the galaxies in the lower mass bin (log($M_*/M_{\odot}$) $<$ 9.8 dex.) The median central $A_V$
is 2.31 and 1.59 mag in the SFB and control samples, respectively, with the sigma (estimated as 1.48 
of the median absolute deviation) of 0.96 and 1.05 mag for the same samples.


\subsection{The Gas Kinematics}

The [SII]$\lambda$6717,6731 line ratio allows us to estimate the ionized
gas density \citep{osterbrock06,perez-montero17} at the centers of SFB galaxies. 
We assume that the gas density decreases exponentially with the distance to the galactic midplane,
see e.g.  \citet{bizyaev17,levy19}.

The ionized gas velocity dispersion at the central outflow area is enhanced in the SFB galaxies. 
Its typical amplitude in our sample is of the order of 100 \kms, and the central and minor axis velocity
dispersion noticeably exceeds that in the rest of the galaxy. 
The gas emission line profiles estimated along the minor axes of the galaxies are systematically 
wider in the SFB galaxies. The maximum minor axes gas velocities are 276 \kms\, in the SFBs, on average,
versus the 165 \kms\, in the regular galaxies from the control sample. These values are estimated
for the gas on the minor axis (within 0.5 kpc from it) at the altitudes above 1 kpc from the galactic
midplane. 
At the same time, this is problematic to estimate the gas outflow
velocities accurately for our highly inclined galaxies: the identified outflows are often shaped as very-narrow 
bi-symmetric structures, with small opening angle. 
Small uncertainties of the opening angle lead to large errors in the outflow velocity.
We assume that the gas escapes from galaxies if its velocity dispersion is increased
due to  
the energy injection from supernovae and young stellar winds. The latter can be 
determined from the star formation surface density $\Sigma_{SFR}$ as
$v \sim \Sigma_{SFR}^\gamma$, where the $\gamma$ can range from 0.18 to 1
depending on the primary reason for the gas turbulence (e.g.
stellar and supernovae feedback, \citet{krumholz16,dib06} or gravitational instability, \citet{krumholz16,lehnert09}).

We estimate the bicone size in the galactic midplane from the MaNGA
\OIII and \Ha images. The typical in-plane bicone diameter is 2 kpc.
We assume that the gas density is an order of magnitude less at the escape altitude than at the midplane.
The mass of the gas at the center $M_{gc}$ is coarsely estimated
under the assumption that all gas is ionized and fills a cylindric volume \citep{heckman90},
whose diameter equals to the size
of the bicone in the galactic midplane, and the height is 1 kpc in all galaxies. 

The equilibrium central velocity dispersion in the galaxies is estimated by \citet{sigma} as
\begin{equation}
\sigma_c = e^{1/2}(0.33\, v_c - 2) \,\,, 
\end{equation}
where the $v_c$ is the maximum circular velocity in the galaxy. 
We assume that the gas starts forming the central outflow when its velocity dispersion exceeds
the $\sigma_c$.

\begin{figure}
\epsscale{1.00}
\plotone{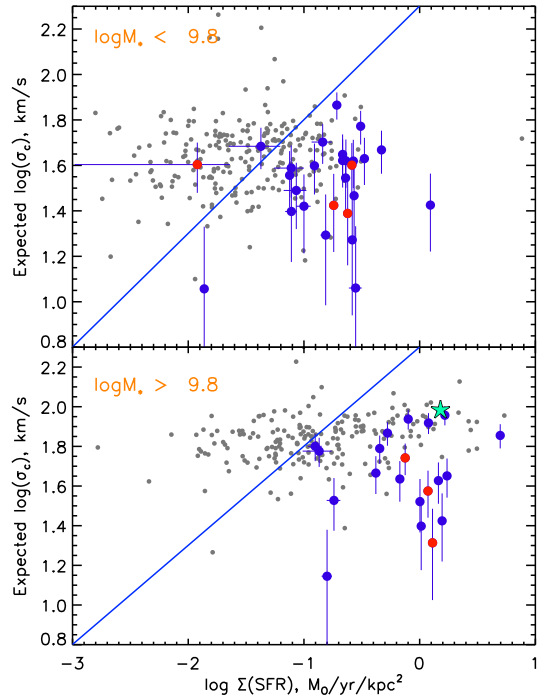}
\caption{The central velocity dispersion in the case of equilibrium ($\sigma_c$, from eq.(1)) in the SFB (blue bullets) and control (grey dots)
galaxies in dependence of the SFR surface density at the center. The red bullets mark the SFB galaxies with 
counter rotation between the gas and stars. The green star designates the galaxy M82. The solid line corresponds to
the case of the equality of $\sigma_c$ to the star formation driven gas velocity dispersion 
$\sigma_g=200\, \Sigma_{SFR}^{1/2}$ at the center (see text). 
The error bars indicate the 2-sigma uncertainty for the SFB objects. 
\label{f3} }
\end{figure}

Large uncertainties in the direct measurement of the gas velocity dispersion 
in highly inclined galaxies make us consider the 
central surface density of the star formation
$\Sigma_{SFR}$, which is is expected to be connected to the  
gas velocity dispersion driven by the star formation
feedback as $\sigma_g \,\sim\, \Sigma_{SFR}^{1/2}$ \citep{krumholz16}.
Figure \ref{f3} compares the the $\Sigma_{SFR}$ (in \Msunyrkpc)
with the gas 
velocity dispersion expected in the disks of galaxies in equilibrium, eq. (1).

The blue line in Figure \ref{f3} corresponds to the case of
$\sigma_g = 200\, \Sigma_{SFR}^{1/2}\,  [\rm{km \, s^{-1}}]$.
We draw the line such that it places all massive SFB galaxies (bottom) and 
all but one low massive SFB galaxies (top) to the right side of the line.
The comparison suggests that the star formation rate in our SFB galaxies
makes the gas central velocity dispersion much higher than required for 
the equilibrium in the galaxies, in contrast with the regular galaxies.

\subsection{The Gas Metallicity}
 
 Similar to the extinction, the strong emission line ratios in elliptical annuli 
corrected for the internal extinction via $A_V$
were used to estimate the gas metallicity. In all cases we required that fluxes in all
emission lines had S/N $>$ 3 and there are no bad reduction flags at the utilized spaxels. 
We consider 3 different metallicity calibrations for the gas metallicity 12+log(O/H):  
N2S2Ha \citep{dopita16}, O2N2 \citep{kewley02} and PG16 \citep{pilyugin16}. 
The first two reveal the same radial distributions in our galaxies,
although with slightly different zero point. We use the N2S2Ha calibrator 
throughout of the paper because it uses 
red emission lines at wavelengths near to each other and thus is 
almost independent of the internal extinction. 
We notice that the PG16 calibrator deminstrates similar radial distributions but shows 
larger uncertainties in the abundances, most probably due to the use of blue and red lines in a mix, 
which allows uncertainties in the $A_V$ affect the estimated metallicity strongly. 
Note that our conclusions based on the same metallicity calibration technique 
for the SFB and control sample should be mostly independent of the chosen metallicity
calibrator's internal accuracy. 

\begin{figure}
\epsscale{1.00}
\plotone{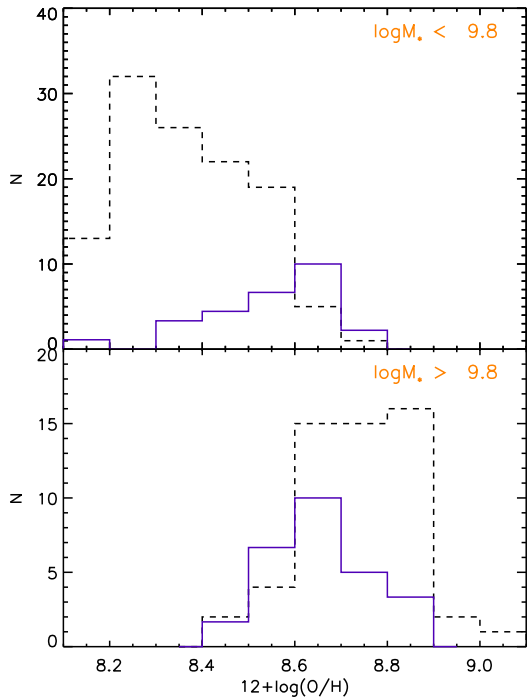}
\caption{The gas metallicity 12+log(O/H) (from the N2S2Ha calibrator) at the central regions 
of the SFB (blue solid line) and control (black dashed line)
galaxies of low (top panel) and high (bottom panel) stellar mass.
\label{f4}}
\end{figure}

Comparison of the radial distributions of the gas metallicity reveals different trends
for the low- and high-mass galaxies. The low mass SFB galaxies have 0.25 dex higher 
central metallicities 12+log(O/H) than the control galaxies of the same mass, 
see Figure \ref{f4}. The comparison for the N2S2Ha calibrator is shown, but the other
considered metallicity calibrators show similar difference. In this case we observe
the outflow of gas enriched by the supernovae explosions. 
The difference in the metallicities and extinction $A_V$ also leads to a more
efficient gas removal from the galactic midplane with the energy and radiation
injected by the star formation process.  
In contrast, massive SFB and control galaxies have similar,
slightly subsolar gas metallicity.

We also notice that the central gas in the low and high mass SFB galaxies
has similar 12+log(O/H) metallicity that varies in narrow with respect
to the control sample ranges, around 8.6 dex. 
Because of the similar star formation density threshold for starting the central outflow in 
all SFB galaxies, one may expect
the narrow metallicity distribution similarity
of the initial, pre-enriched gas in them. In turn, it points towards the external 
origin of the gas, which can infall from larger galaxies in groups, 
like in the case of M82, where a larger galaxy M81 supplies the gas 
to the M82 biconical outflow \citep{lehnert99}. 

\subsection{Stellar Population}

The main MaNGA data reduction pipeline \citep{law16} reports all Lick indices \citep{worthey97}, 
their uncertainties, and data quality flags. 
We study the radial distributions of age and metallicity indicators determined
from the absorption spectra features averaged over the 1 kpc wide elliptical annuli or
zones same way as described above. We took into account only those spaxels where the indices
have S/N $>$ 3 and the data reduction flags are good. 
We use the \ddd index \citep{gorgas99} as the average stellar population age indicator. 
For the iron abundance we use the $<{\rm Fe}> \,=\, {\rm Fe}5270 + {\rm Fe}5335$ combination. 
The [$\alpha$/Fe] is estimated using the strongest Mg feature Mgb as 
$[\alpha/{\rm Fe}] \,=\, -1.030 + 1.016\, X - 0.141\, X^2$, where $X = {\rm Mg\,b}/<{\rm Fe}>$, which is a fit to 
the relation by \citet{thomas02}.

\begin{figure}
\epsscale{1.00}
\plotone{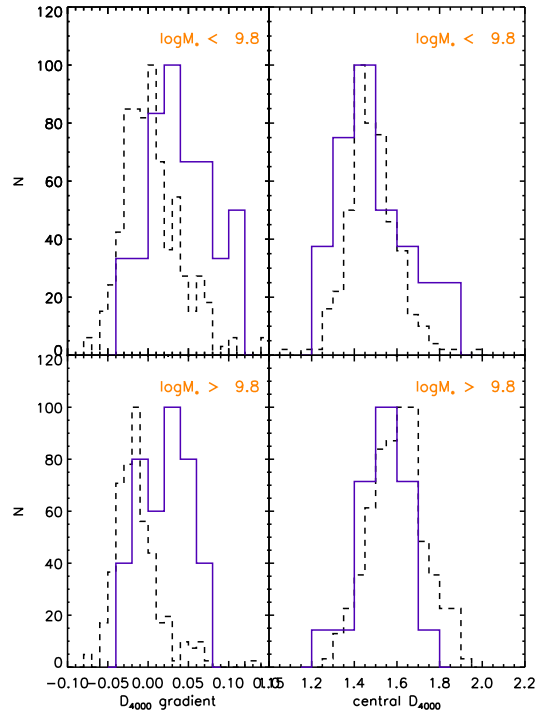}
\caption{The central values and gradients of the \ddd index for the 
SFB (blue solid) and control (black dashed) galaxies of low (top panel) and high (bottom panel) stellar mass.
The peaks of all distributions are normalized by 100. 
\label{f5}}
\end{figure}

Figure \ref{f5} contrasts the central value and
gradient of the \ddd index in the SFB and control galaxies. The \ddd sensitive to
the age of stellar population is not different at the centers, while the \ddd radial gradient
is high positive in SFB hosts, in a contrast with mostly zero or negative
gradients in the regular galaxies.  
Since \ddd traces the efficient age of underlying stellar population, it suggests 
a sharp concentration of significant population of young stars in the SFBs. 
Such a peak of young objects at the 
centers points towards a prior infall of building material (gas) 
to the central region. 
 
\begin{figure}
\epsscale{1.00}
\plotone{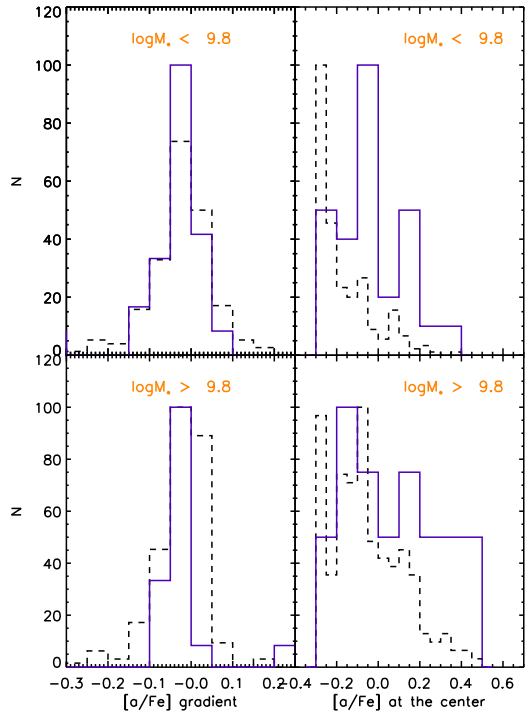}
\caption{The central values and gradients of the \aFe for the 
SFB (blue solid) and control (black dashed) galaxies of 
the low and high mass ranges (top and bottom panel, respectively.)
The peaks of all distributions are normalized by 100. 
\label{f6}}
\end{figure}

Comparison of effective metallicity of the stellar population 
from the Fe5270 and Fe5335 indices do not
show a significant difference for the SFB galaxies. At the same time, the \aFe
estimated from a combination of Mgb and Fe indices reveals
high fraction of alpha-enhanced centers of SFB galaxies, see Figure \ref{f6}. The latter
fact indicates a relatively short age (much less than 1 Gyr) of the central starburst 
due to a time lag of the SNe Ia \citep{mcwilliam97}.

\section{Discussion}

Contrasting the large sample of the SBF and regular galaxies allows  us to 
notice differences in observational properties of these galaxies and to make
conclusions about the distinctive features of the SFB galaxies. 
 
 {\it The driver of the central outflows.} 
The enhanced SF surface density at the center 
is distinctive features of the galaxies 
with biconical outflows (Figure \ref{f2a}). 
We compare the SF driven velocity dispersion in the ionized gas
with the equilibrium velocity dispersion in Figure \ref{f3} and conclude that 
the gas must expand inn the direction perpendicular to the galactic midplane
and finally escape to high galactic altitudes. 
A good illustration of the rapid and centrally localized star burst in the SFBs
comes from systematically high radial gradient of \ddd index, see Figure \ref{f5}, which suggests
a large difference of stellar age along the radius in the SFB galaxies: they have
much younger stellar population at the center than at the outer regions.
At the same time, the centers of SFBs demonstrate enhanced abundance
of alpha-elements traced by the [Mg/Fe] indicator. These facts suggest
that the rapid star formation started much less than 1 Gyr ago, and that 
since that time it has built significant amount of new stars sufficient to 
make the photometric profiles steeper and the overall effective radii shorter. 

{\it Important conditions for creating and maintaining the SFBs.}
We notice systematically higher $A_V$ and the gas metallicity in the low mass SFB 
galaxies with respect to the regular ones (Figure \ref{f4}).
The higher gas  
extinction help couple the radiation with gas, and to push the gas from
the galaxy more efficiently (see \citet{zhang18} and references therein). 
These factors also may be responsible for the star formation 
density threshold decreasing \citep{heckman02} responsible for the outflow emergence 
below the 0.1\Msunyr in small galaxies, see Figure \ref{f3}.
We consider the ratio of the estimated and the equilibrium gas velocity dispersions
with an addition of empirical dependence on the metallicity and extinction, 
$f ~=~ log [ \Sigma_{SFR}^{1/2} \, Z ~/~ \sigma_c ]$, where the SFR surface
density $\Sigma_{SFR}$ is 
measured at the center from the \Ha luminosity,
$Z = \tau_V \, 40*10^{(O/H)}$, 
$\tau_V$ is the optical depth from the extinction $A_V$, 
$\sigma_c$ is the equilibrium velocity dispersion at the center of
galaxy from eq. (1), and the gas velocity dispersion is assumed to be driven by the 
star formation feedback \citep{krumholz16}, for which the $\sigma \,\sim\, \Sigma_{SFR}^{1/2} $. 
Figure \ref{f6} (left) shows the comparison of the ratio $f$ with
the effective radii $R_e$ for the galaxies in our samples. 
By introducing the empirical value $f$ we attempt to find a combination
of parameters suitable for selecting the SFB galaxies.
The SFB galaxies are rather well separated from the regular ones
in Figure \ref{f6} (left), which suggests the importance of
of the factors contributing to the $f$ for maintaining the SFBs. Figure \ref{f6} (right) shows the distribution 
of the figure of merit $F$ introduced as $F = f - 2 \log\, R_e$. Using the 
$F$, we can efficiently separate the SFB galaxies from the regular ones.

\begin{figure}
\epsscale{1.00}
\plotone{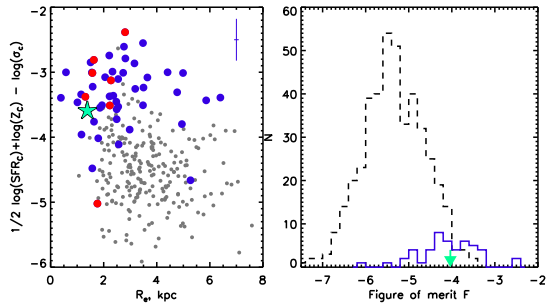}
\caption{Left: the ratio of the star formation driven to the equilibrium velocity dispersions versus the effective radius $R_e$.
The former gas velocity dispersion is estimated in arbitrary units as $\Sigma_{SFR}^{1/2}\, Z_c$, 
where $Z_c = 40\cdot 10^{(O/H)}\, \tau_V$ (see text). The blue bullets denote the SFB galaxies, the red bullets highlight the 
SFB galaxies with a counter-rotation between the gas and stars. The grey dots show the control sample.
The error bars at the top right corner indicate the typical uncertainty.
\\
Right: the figure of merit F =$log[\Sigma_{SFR}^{1/2}\, Z_c] - log(\sigma_c) - 2\,logR_e$ in the SFB (blue solid line)
and control (grey dashed line) galaxies. The green arrow designates the galaxy M82. 
\label{f7}}
\end{figure}

{\it The central concentration of the outflows.}
If the gas outflow 
condition ($\sigma_g \,>\, \sigma_c$, Figure \ref{f3}) were fulfilled in the whole galaxy, the gas would
escape from all regions of the galactic body. The central localization of the outflow is
caused by the concentration of star formation to the center. 
The SFB galaxies have systematically shorter $R_e$ than the 
control galaxies. In turn, the ignition of the powerful central starburst
in the SFB galaxies is impossible without driving large amount of gas at the center, sufficient
for creating significant amount of your stars, which creates central concentration of 
young stellar population.    

{\it The life time of the SFBs.}
We estimate the outflow rate $\dot M_{out}$ in our SFB galaxies using the central 
electron density as
$\dot M_{out} \,=\,  1.4 \pi\, m_p\, v_{out}\, n_e \, r_c^2$,
where $m_p$ is the mass of proton, $n_e$ is the electron density, $r_c$ is the 
bicone radius (see \S3.2), $v_{out}$ is estimated as 
$v_{out} = 140 \Sigma_{SFR}^{1/2}$ \citep{krumholz16,yu19}, and the factor 1.4 accounts for 
the contribution of heavy elements to the total gas mass. The electron density widely ranges
 from 7 to 286 cm$^{-3}$ across the SFB sample, with median value of 70  cm$^{-3}$.

\begin{figure}
\epsscale{1.00}
\plotone{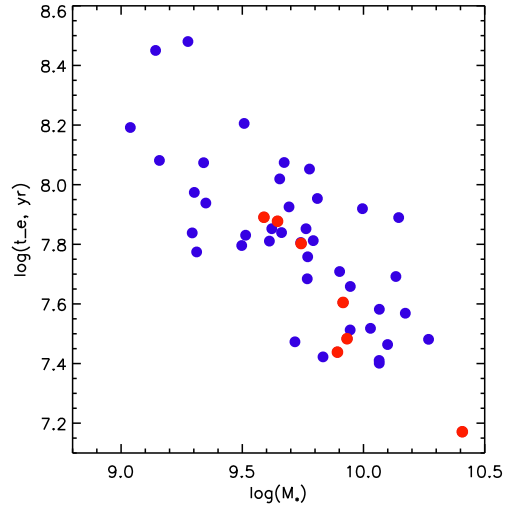}
\caption{The estimated outflow exhaust time in the main SFB sample (blue bullets) and in
seven SFB galaxies with a gas-stars counter-rotation (red bullets) anti-correlates with the galactic 
stellar mass. 
\label{f8}}
\end{figure}

The median mass loss rate $\dot M_{out}$ is 14.3 \Msunyr for our SFB sample.
At the same time we assess the ionized gas mass $M_{gc}$ at the center
from the volume and density
and conclude that the outflow cannot last more than a few hundred Myr: 
the exhaust time $t_e = M_{gc}/ \dot M_{out}$ in our SBF sample ranges from
20 to 300 Myr (the median time $t_e$ is 65 Myr). The $t_e$ anti-correlates with the galactic stellar 
mass: it is long in our low mass SFB galaxies and short
in the massive galaxies, see  Figure \ref{f8}. The correlation agrees with a natural assumption that 
the gravitational potential of massive galaxies attracts more circumgalactic 
gas than for small galaxies. More external gas falls to the central area between 
the SFB active cycles and supports more intensive star formation
during the active cycles in the large galaxies. 
It is worth noting that the SFB galaxies with counter-rotation
(red bullets in Figure \ref{f8}) indicate shorter exhaust time for the outflow, on average, with respect
to the other SFB objects. In those cases the external gas - internal gas interaction
can provide more efficient transportation of the gas to the central regions due to a more
efficient gas momentum loss, 
which should shorten the SFB recharging time and the whole duty cycle. 

Note that the star formation rate $\dot M_{SF}$ at the center is an order of magnitude
less than the outflow rate ($\dot M_{out} \gg \dot M_{SF}$), which makes the biconical outflow the 
principal regulator of the gas mass balance at the central regions. The typical 
gas infall rate in regular galaxies is comparable to the star formation rate \citep{chiappini09}, 
so a steady infall with a constant rate cannot compensate the gas outflow in 
the SFB regime. If the gas were supplied in a slow steady infall only, the star formation 
bicones would show some 1.5 Gyr long "recharging" cycles of the bicone inactivity in 
large galaxies, and much longer than that in dwarfs. The latter suggests that the SFB
activity should be triggered by a bulk accretion of gas (e.g. due to a minor merging).

{\it The triggers for the SFB outflows.} 
The central starburst started
in our SFB galaxies less than a few hundred Myr ago, which makes it a 
temporary phenomenon in galaxies, and requires a triggering event. 
The gas necessary for feeding the starburst
can be of the external or internal origin \citep{lehnert96b}. Neither of the SFB galaxies
in our sample shows traces of recent or ongoing minor 
merging or interaction,
which excludes recent major mergers.

The accretion of gas from distant companion galaxies or from small, gas-rich satellites
to the central area of our SFB galaxies would provide gas supply 
necessary to start the starburst, and also would explain the gas metallicity distribution
in the SFB galaxies.
A high fraction of objects with a counter rotation between the gas and stars
among the SFB galaxies (15\%, with no counter-rotation in the control sample) 
supports the hypothesis that the gas accretion from satellites can often trigger the SFB ignition. 
We find that 56\% of our SFB galaxies have large or small companions,
or are group members, although we do not see traces of interaction on
deep MzLS+BASS images\footnote{http://legacysurvey.org/pubs/}. The other 17\% of the sample
indicate noticeable asymmetry of external isophotes, which suggests that a
minor merging took place in the past. 
This mechanism is also suggested for low mass galaxies with counter-rotation between
gas and stars, where the central star formation rate is high and young population 
dominates \citep{chen16}.

An alternative mechanism, a rapid internal redistribution of
gas in galaxies, e.g. a result of bar-driven gas inflows, 
may be responsible for driving a large amount of gas to the 
central regions and for igniting the star formation burst.
In this case the internal reasons should be able to increase the 
rate of transferring the gas to the central region, which may
also be caused by enhanced external gas infall (e.g. as a result of 
an accretion event to the outer regions of galaxy). 
Both large positive gradients of \ddd and  age and in the \aFe indicates 
the concentration of recently formed young stars to the centers of SFB galaxies. 
In a combination with large velocity dispersion of gas and stars there it 
suggests that we may observe the growth of young pseudo-bulges \citep{lackner13},
which can be also connected to the secular processes in galactic 
disks \citep{kormendy04,athanassoula04}.

While the central outflows remove the gas from the sites of active star formation,
we do not see an evidence that the gas leaves the galaxies forever, as it has
been noticed by \citet{oppenheimer06,oppenheimer10,leroy15,emonts17,rupke18}. 
Instead, the gas expelled from the 
central regions enriches the intergalactic medium and may return back to
refuel the starburst-galactic outflow cycle. The properties of many galaxies in
our sample resemble those of a well studied nuclear starburst galaxy M~82, 
which helps guess the star formation
rates, outflow rates, and the warm gas budget at the centers of our galaxies. 
Our estimates show that the outflow rates exceed the inflow rates by the order 
of magnitude. We also expect
that our estimates of the gas outflow rates set only the lower limit because the 
hot gas in the outflow entrains the warm and cold gas components (see e.g. \citet{rupke18,zhang18}),
while our observations are sensitive to the warm component of the gas media only. 
The intensive mass loss with respect to the gas replenishing rate suggests
a temporary nature of the SFB and points at the necessity of prolonged "gas recharging" time. 
Our criteria for the SFB galaxies selection will help select and study more clear
cases of face-on SFB galaxies, in which the kinematics of the outflows can
be studied directly from ongoing spectroscopic observations, and the gas mass circulation
rate can be estimated in a more straightforward way.


\acknowledgments
D.B. is partly supported by RSCF grant 19-12-00145.
Y. C. acknowledges support from the National Key R\&D Program of China (No. 
2017YFA0402700), the National Natural Science Foundation of China (NSFC
grants 11573013, 11733002)
Y.S.  acknowledge support
from the National Key R\&D Program of China (No.  2018YFA0404502) and the
National Natural Science Foundation of China (NSFC grants 11733002 and
11773013). RR thanks to CNPq, FAPERGS and CAPES for financial support.

We thank the anonymous referee for valuable feedback that improved the paper.

SDSS-IV acknowledges support and resources from the Center for
High-Performance Computing at the University of Utah.  The SDSS web site
is www.sdss.org.

SDSS-IV is managed by the Astrophysical Research Consortium for the
Participating Institutions of the SDSS Collaboration including the
Brazilian Participation Group, the Carnegie Institution for Science,
Carnegie Mellon University, the Chilean Participation Group, the French
Participation Group, Harvard-Smithsonian Center for Astrophysics,
Instituto de Astrof\'isica de Canarias, The Johns Hopkins University,
Kavli Institute for the Physics and Mathematics of the Universe (IPMU) /
University of Tokyo, Lawrence Berkeley National Laboratory, Leibniz
Institut f\"ur Astrophysik Potsdam (AIP), Max-Planck-Institut f\"ur
Astronomie (MPIA Heidelberg), Max-Planck-Institut f\"ur Astrophysik (MPA
Garching), Max-Planck-Institut f\"ur Extraterrestrische Physik (MPE),
National Astronomical Observatory of China, New Mexico State University,
New York University, University of Notre Dame, Observatrio Nacional /
MCTI, The Ohio State University, Pennsylvania State University, Shanghai
Astronomical Observatory, United Kingdom Participation Group, Universidad
Nacional Aut\'onoma de M\'exico, University of Arizona, University of
Colorado Boulder, University of Oxford, University of Portsmouth,
University of Utah, University of Virginia, University of Washington,
University of Wisconsin, Vanderbilt University, and Yale University.

{}

\end{document}